\documentclass[aps,prl,10pt,superscriptaddress,twocolumn,preprintnumbers,amsmath,amssymb]{revtex4-2}

\usepackage{mathtools}
\usepackage{slashed}
\usepackage{tensor}
\usepackage{bbold}  
\usepackage{amscd}
\usepackage{color}
\usepackage{slashed}
\usepackage[normalem]{ulem}

\usepackage{calrsfs}  
\DeclareMathAlphabet{\pazocal}{OMS}{zplm}{m}{n}

\allowdisplaybreaks

\newcommand{\tr}{\mathrm{tr}\,}

\newcommand{\bv}[1]{\boldsymbol{#1}}

\usepackage[dvipsnames]{xcolor}

\definecolor{brickred}{rgb}{0.8, 0.25, 0.33}
\newcommand\myshade{85}
\colorlet{mylinkcolor}{BrickRed}
\colorlet{mycitecolor}{NavyBlue}
\colorlet{myurlcolor}{Aquamarine}

\usepackage{hyperref}
\hypersetup{
  linkcolor  = mylinkcolor!\myshade!black,
  citecolor  = mycitecolor!\myshade!black,
  urlcolor   = myurlcolor!\myshade!black,
  colorlinks = true,
}

\begin{document}

\title{Curvature-Controlled Infrared Regularization of Crystalline Membranes}

\author{Pablo A. Morales}
\email{pablo$_$morales@araya.org}
\affiliation{Research Division, Araya Inc., Tokyo 101-0025, Japan}
\affiliation{Centre for Complexity Science, Imperial College London, London SW7 2AZ, UK}

\author{Pavel Castro-Villarreal}
\email{pcastrov@unach.mx}
\affiliation{Facultad de Ciencias en Física y Matemáticas, Benemérita Universidad Autónoma de Chiapas, Carretera Emiliano Zapata, Km. 8, Rancho San Francisco, C. P. 29050, Tuxtla Gutiérrez, Chiapas, México}

\begin{abstract}
We show that negative Gaussian curvature regularizes the infrared sector of crystalline membranes. In a covariant formulation of embedded elasticity, the Green strain contains a symmetry-required linear coupling between the normal phonon field and extrinsic curvature. Integrating out the in-plane phonons converts this coupling into a finite quadratic contribution to the inverse flexural response. Anomalous roughening is thereby replaced by curvature-controlled saturation, and the mechanism survives on minimal hyperbolic patches. Hyperbolic geometry preempts anomalous elasticity before the flat infrared regime is reached, implying the absence of a crumpling phase at the harmonic level. The same Gaussian-order coupling admits sound propagation in the infrared.
\end{abstract}

\maketitle

\paragraph{Introduction.}
The flat phase of a thermalized crystalline membrane is anomalous at long wavelengths. Integrating out in-plane phonons about a flat reference plane generates the nonlocal Nelson--Peliti interaction~\cite{nelson1987fluctuations}, which drives the infrared sector to a strongly coupled fixed point: the bending rigidity diverges as $\kappa(q) \sim q^{-\eta_A}$, the in-plane Young's modulus vanishes as $Y(q) \sim q^{\eta_u}$ and the two exponents are tied by a Ward identity $2\eta_A + \eta_u =2$ \cite{aronovitz1988fluctuations, bowick1996flat}. Read literally, this fixed point describes a membrane that stiffens without bound at long wavelengths and possesses no propagating flexural acoustic branch in the asymptotic regime. Furthermore, at harmonic order the mean-square fluctuation of the surface normals diverges logarithmically as $L \to \infty$, signaling a tendency to crumple~\cite{nelson2004statistical}.
The experimental situation is in direct tension with these predictions; a well-defined flexural acoustic mode is observed in freely suspended graphene at wavevectors well inside the window where anomalous scaling is supposed to dominate~\cite{PhysRevLett.127.266102, Tomterud2025}. The long-wavelength rigidity of real two-dimensional crystals is not divergent,  and the flexural sector carries sound. While intrinsic corrugations in 2D materials are experimentally well established~\cite{meyer2007structure}, the crumpling transition originally predicted for tethered membranes~\cite{Kantor_1987} has not yet been conclusively observed.

The anomalous flat phase has been the subject of sustained theoretical effort~\cite{PhysRevE.105.L012603, Metayer_2024, PhysRevE.103.053002}, self-consistent screening of polymerized membranes~\cite{PhysRevLett.69.1209}, Monte Carlo studies of the discretized~\cite{Bowick:2001bb} and atomistic~\cite{Fasolino2007} models, couplings to Dirac fermions in conducting crystals~\cite{Gazit2009, Guinea_2014} and most recently non-perturbative lattice-dynamics calculations within the stochastic self-consistent harmonic approximation~\cite{aseginolaza2024bending}. The last identifies that invariance under rotations of the surface embedded in $\mathbb{R}^3$, not of its internal coordinates, requires retaining the in-plane nonlinearity $\partial_a u^c \partial_b u_c$ in the Green strain, a term second order in the phonon fields, routinely discarded by the standard small-fluctuation expansion.

Beneath all of this lies the often overlooked assumption that the reference geometry is flat. This sits in tension with the very mechanism by which 2D crystals exist. The Mermin–Wagner theorem forbids long-range order in 2D harmonic systems at finite temperature~\cite{Mermin1966}; atomically stable thin crystals are sustained by the substantial out-of-plane fluctuations of their flexural sector, and the configuration about which those fluctuations are expanded cannot itself be flat. Empirically, freely suspended graphene, hexagonal boron nitride, transition-metal dichalcogenides, and related atomically thin crystals stabilize through static nanometric corrugation~\cite{meyer2007structure, Deng2016, chen2019wrinkling}. Once curvature is present, the embedding geometry must be expanded covariantly, and the symmetries of the Green strain reimposed order by order in the phonon fields. Flattening the reference geometry at the outset discards the leading geometric content of the embedded theory.

In this Letter, we show that the covariant treatment alters the infrared problem already at Gaussian order. The exact Green strain on a curved surface contains a symmetry-required coupling, linear in the phonon fields between the normal displacement and the extrinsic curvature, at the same order as the harmonic in-plane strain itself, and therefore retained by any consistent small-fluctuation truncation. Integrating out the in-plane phonons converts this coupling into a contribution to the inverse flexural propagator that, on surfaces of negative Gaussian curvature, is finite and strictly positive at zero wavevector. The long-wavelength response is regular so the flexural structure factor saturates, the membrane roughness converges to a finite plateau rather than continuing its anomalous growth, and the flexural mode acquires a finite long-wavelength frequency fixed by the local curvature and the elastic moduli, implying the lack of a crumpling phase. The mechanism is intrinsic to negative Gaussian curvature; 
hyperbolic geometry therefore preempts anomalous elasticity, and the flat-phase infrared never sets in.

\paragraph{Covariant elasticity.} 
The reference configuration of a suspended 2D crystal is not flat. Its mechanical stability is sustained by corrugations~\cite{Deng2016,chen2019wrinkling}, directly observed as nanometric ripples in free-standing graphene~\cite{Fasolino2007}. We therefore take the reference to be a smooth surface $\Sigma$ embedded in $\mathbb{R}^3$ through ${\bv X}(x^{a})$, where $\{x^{a}\}$ are local coordinates with $a=1,2$. The tangent basis and unit normal are denoted by $\{{\bv e}_a\}$ and ${\bv n}$, respectively, with induced metric $g_{ab}={\bv e}_a\cdot {\bv e}_b$ and extrinsic curvature $K_{ab}={\bv e}_{a}\cdot\partial_{b}{\bv n}$. A deformation of the surface is decomposed into tangential and normal parts, ${\bv X}'(x)={\bv X}(x)+ u^a(x){\bv e}_a(x)+\eta(x){\bv n}(x)$, where $u^{a}(x)$ and $\eta(x)$ are the in-plane and out-of-plane elastic phonon fields, respectively \cite{PhysRevE.103.053002}. The deformed tangent vector is then
\begin{equation}
    {\bv e}'_a = \left(\delta_a^b+\nabla_a u^b+\eta K_a{}^b\right){\bv e}_b + \left(\nabla_a\eta-K_{ab}u^b\right){\bv n},
\end{equation}
where $\nabla_{a}$ is the covariant derivative compatible with $g_{ab}$.  Thus the nonlinear strain is the geometrical Green strain tensor $E=E_{ab}d x^{a}\otimes dx^{b}$, with components $E_{ab}=\frac12(g'_{ab}-g_{ab})$ given by~\cite{ciarlet2005introduction}
\begin{align}
    E_{ab} &= \nabla_{(a}u_{b)}+\eta K_{ab} \nonumber\\
    &\hphantom{=}\;+ \frac12 \left(\nabla_a u^c+\eta K_a{}^c\right)
    \left(\nabla_b u_c+\eta K_{bc}\right) \nonumber\\
    &\hphantom{=}\;+ \frac12 \left(\nabla_a\eta-K_{ac}u^c\right)
    \left(\nabla_b\eta-K_{bd}u^d\right).
    \label{eq:exact_strain}
\end{align}
The strain~\eqref{eq:exact_strain} represents the covariant form of the components of the Green tensor $E$ and is invariant under diffeomorphisms and rigid motions of the embedded surface in $\mathbb{R}^3$~\cite{supp}. In the flat limit $g_{ab} = \delta_{ab}$ and $K_{ab}= 0$, the strain tensor components $ E_{ab} $ simplify to the flat membrane strain \cite{nelson2004statistical}. The coupling between the normal phonon and the extrinsic curvature in~\eqref{eq:exact_strain} has no flat counterpart.

The elastic energy of the deformed membrane from a curved surface reference consists of two terms $H=H_{\rm st}+H_{\rm bend}$, with $H_{\rm st}$ the stretching energy given by
\begin{equation}
    H_{\rm st}[{\bv X}, u^{a}, \eta] = \frac12\int_\Sigma dA\, C^{abcd}E_{ab}E_{cd},
    \label{eq:H_st}
\end{equation}
where $C^{abcd} = \lambda g^{ab}g^{cd} + \mu (g^{ac}g^{bd}+g^{ad}g^{bc})$ is the isotropic 2D elastic tensor and $\lambda$, $\mu$ the Lamé constants. The bending energy is given by the Helfrich--Canham term 
\begin{eqnarray}
    H_{\rm bend}[{\bv X}, u^{a}, \eta] =\int_{\Sigma'} dA^{\prime}\,\left(\frac{\kappa}{2}K'^2+\sigma\right),
\end{eqnarray}
where all primed geometric quantities are evaluated on the deformed
surface using the embeddings ${\bv X}^{\prime}(x)$, $\kappa$ is the bending rigidity modulus and $\sigma$ the surface tension. The total Hamiltonian $H$ retains the symmetries of $E$.

\paragraph{Normal-mode effective theory.} Thermal fluctuations of the membrane are governed by the partition function $\mathbb{Z}=\int \mathcal{D}\eta\mathcal{D}u^{a}e^{-\beta H}$, with $H$ nonlinear in the phonon fields. The standard treatment rests on two simplifications~\cite{nelson1987fluctuations, aronovitz1988fluctuations, PhysRevLett.69.1209, Gazit2009, Guinea_2014, aseginolaza2024bending, Kokovin2024}: (i) the reference equilibrium is taken flat, and (ii) the in-plane anharmonic term $\partial_i u^k \partial_j u_k$ is discarded, breaking the embedded SO($3$) and identified in~\cite{aseginolaza2024bending} as the term governing the anomalous flat fixed point. Under these simplifications, the long-wavelength bending rigidity diverges. We retain the reference curvature and show that it controls the infrared already at Gaussian order, unlike the anharmonic contribution; the mechanism below is
therefore independent of the nonlinearity. 

The bending contribution decouples from the in-plane integration. A tangential deformation $ \delta_u {\bv X} = u^a {\bv e}_a $ acts as a bulk reparametrization, so $\delta_u H_{\rm bend} = \int_{\Sigma} dA\,\nabla_a\left(u^a {\cal F}_{\rm bend}\right)$ reduces to a boundary term that vanishes for closed surfaces or boundary conditions removing it~\cite{supp}. The expansion of $H_{\rm bend}$ around $\Sigma$ is 
\begin{eqnarray}
    H_{\rm bend}[{\bv X}, \eta] = H_{\rm HC}[{\bv X}] + \frac{1}{2} \int dA\, \eta {\cal L}_{\rm HC} \eta + \cdots,
    \label{eq:Helfrich-Canham-low}
\end{eqnarray}
where $ H_{\rm HC}[{\bv X}] $ is the bending energy with no surface deformation and $ {\cal L}_{\rm HC} $ is a fourth-order differential operator on $\Sigma$~\cite{PhysRevB.110.195430}. The partition function factorizes,
\begin{eqnarray}
    \mathbb{Z} = \pazocal{N} \int \mathcal{D}\eta \,e^{ -\frac{\beta}{2} \int dA\, \eta {\cal L}_{\rm HC} \eta} \int \mathcal{D}u^{a} e^{-\beta H_{\rm st}[{\bv X}, u^{a}, \eta]},
\label{eq:partition-function}
\end{eqnarray}
so the in-plane phonons decouple from the flexural ones; $\pazocal{N}$ is a constant that depends on the curved surface reference. Even if rotational symmetry is broken, considering small fluctuations in the phonon fields is now plausible as a physical approximation.

To isolate the leading curvature-induced normal-mode kernel, we expand~\eqref{eq:exact_strain} to first order in the phonon fields,
\begin{equation}
    E^{(1)}_{ab}
    = (Du)_{ab} + \eta K_{ab},
    \qquad (Du)_{ab}\coloneqq \nabla_{(a}u_{b)} .
    \label{eq:linear_strain}
\end{equation}
This Gaussian phonon truncation is sufficient for the quadratic kernel in $\eta$. The nonlinear pieces in~\eqref{eq:exact_strain} generate cubic and quartic phonon vertices whose flat-limit contributions drive the Nelson--Peliti nonlocal interaction and renormalize the elastic parameters~\cite{nelson1987fluctuations}. 
On the curved reference they are subleading corrections to the Gaussian kernel obtained below. The decisive feature is that the linear coupling $\eta K_{ab}$ already enters the quadratic Hamiltonian.
Substituting~\eqref{eq:linear_strain} into~\eqref{eq:H_st} gives 
\begin{align}
    H_{\rm st}^{(2)}
    &=\int dA\, C^{abcd} \bigg[ \frac12  (Du)_{cd} (Du)_{ab} \nonumber\\ 
    &\hphantom{=\int dA\, C^{abcd}}\;\; +\eta (Du)_{ab} K_{cd} + \frac12 \eta^2 K_{ab}K_{cd} \bigg].
\end{align}
After integrating by parts, the pure in-plane part can be written as $\frac12\int dA\, u_b\mathcal{D}^b{}_c u^c$, where the covariant phonon operator is
\begin{equation}
    \mathcal{D}^b{}_c = -\mu \Delta \delta^b_c - (\lambda+\mu)\nabla^b\nabla_c - \mu R^b{}_c .
\end{equation}
Here $R^b{}_c$ is the Ricci tensor of $\Sigma$. The mixed term defines a curvature-induced source for the in-plane field,
\begin{equation}
    J^b(\eta) = -\nabla_a
    \left[ \eta ( \lambda g^{ab}K+2\mu K^{ab} ) \right],
    \label{eq:source}
\end{equation}
where $K=g^{ab}K_{ab}$. Carrying out the in-plane phonon $u^{a}(x)$ integration~\eqref{eq:partition-function}, the effective quadratic stretching energy for the normal displacement,
\begin{align}
    H_{\rm st,eff}^{(2)}[\eta]
    &= \frac12\int dA\,\eta^2
    (\lambda K^2+2\mu K_{ab}K^{ab} ) \nonumber\\
    &\hphantom{=}\;- \frac12 \int dA_x dA_y\,
    J_b(x)G^b{}_c(x,y)J^c(y),
    \label{eq:Heff_quad_general}
\end{align}
where $G^b{}_c(x,y)$ is the Green function corresponding to the operator $\mathcal{D}^b{}_c$.  
As shown in~\eqref{eq:Heff_quad_general}, a curved reference surface differs qualitatively from a flat one already at quadratic order because a normal displacement changes the metric linearly through $\eta K_{ab}$; the eliminated in-plane phonons generate a normal-mode kernel before any quartic height interaction is revealed.

To extract the infrared structure of the covariant quadratic kernel we assume the equilibrium configuration as an open, extended surface whose
static corrugation intercalates regions of positive and negative Gaussian curvature. The infrared behavior is approximated in each local patch of the surface, as shown below, this is sufficient to generate a mass term in the flexural sector. 
We work in Riemann normal coordinates (RNC) centered at $p\in\Sigma$. We keep the leading constant-coefficient part of the local RNC expansion, evaluating $K_{ab}$ and $R_{ab}$ at $p$ and replacing the covariant derivative, $\nabla_{a}$, acting on the local mode by $iq_a$. 
All geometric quantities in this section are evaluated at $p$ unless otherwise stated. The truncation is the least favorable setting in which to expose a curvature effect. Gradient corrections such as $\nabla K_{ab}$ and $\nabla R_{ab}$ enter the kernel only through insertions against an in-plane propagator that is already gapped; they are therefore infrared-finite. Any regularization persisting at this order is thus a lower bound on the full effect.

In two dimensions the Ricci tensor satisfies $R_{ab}=K_G g_{ab}$
where $K_G$ is the Gaussian curvature. The local momentum-space form of the in-plane phonon operator is therefore ${\cal D}^{a}{}_b(q) = \mu(q^2-K_G)\delta^{a}_b + (\lambda+\mu)q^a q_b$. Introducing the longitudinal and transverse projectors
\begin{equation}
    P^L_{ab}(q)=\frac{q_aq_b}{q^2},
    \qquad
    P^T_{ab}(q)=\delta_{ab}-\frac{q_aq_b}{q^2},
\end{equation}
it becomes
\begin{equation}
    {\cal D}_{ab}(q) = \mu(q^2-K_G)P^{T}_{ab}
    + \left[(2\mu+\lambda)q^2-\mu K_G\right]P^L_{ab}.
\end{equation}
Thus, the local phonon Green function in momentum space is
\begin{equation}
    G_{ab}(q) = \frac{P^T_{ab}(q)}{\mu(q^2-K_G)}
    + \frac{P^L_{ab}(q)}{(2\mu+\lambda)q^2-\mu K_G}.
\end{equation}
This is the curved analog of the usual flat longitudinal-transverse phonon propagator. The crucial difference is that curvature enters the denominator of both polarization channels. We next resolve the curvature source in the same local frame. Setting the orthonormal principal frame of the second fundamental form as $K_{ab}={\rm diag}(k_1,k_2)$, where $K=\tr (K_{ab})$ and $K_G = \det(K_{ab})$. For a wavevector ${\bv q}=q {\bv n}$, with $\bv{n}=(\cos\theta,\sin\theta)$, and $\bv{t}=(-\sin\theta,\cos\theta)$,
we can define the normal curvature and geodesic torsion in the direction of propagation, $k_n=K_{ab}n^a n^b$ and $\tau_g=K_{ab}n^a t^b$. Neglecting derivatives of $K_{ab}$ at this order, the source~\eqref{eq:source} has longitudinal and transverse components
\begin{align}
    J_L &\equiv n^bJ_b
    = -iq(\lambda K+2\mu k_n)\eta({\bv q}), \\
    J_T &\equiv t^bJ_b 
    = -2i\mu q\tau_g \eta({\bv q}).
\end{align}
Substitution into the effective phonon energy gives $H_{\rm st,eff}^{(2)}[\eta]= \frac12\int \frac{d^{2}q}{(2\pi)^{2}} \mathcal{E}_{\rm st}({\bv q}) |\eta({\bv q})|^2$, with mode energies 
\begin{align}
   \mathcal{E}_{\rm st}({\bv q})
    &= (\lambda+2\mu) K^2-4\mu K_{G} \nonumber\\
    &- q^2
    \left[
        \frac{4\mu \tau_g^2 (\theta)}
        {q^2-K_G}
        + \frac{[\lambda K+2\mu k_n(\theta)]^2}
        {(2\mu+\lambda)q^2-\mu K_G}
    \right],
    \label{eq:local_kernel_general}
\end{align}
where we have employed the Gauss-Codazzi $K_{G}=\frac{1}{2}(K^{2}-K_{ab}K_{ab})$. This kernel is the stretching contribution to the flexural modes.

\paragraph{Hyperbolic infrared regularization.}
For a hyperbolic patch, $K_G=-|K_G|$. Hence, the denominators in~\eqref{eq:local_kernel_general} are $q^2+|K_G|$ and $(2\mu+\lambda)q^2+\mu |K_G|$ respectively, both remain finite at $q=0$. The local in-plane Green function therefore has no local infrared pole. 
Moreover, the phonon-mediated subtraction in~\eqref{eq:local_kernel_general}  begins at order $q^2$ rather than renormalizing the constant part of the normal kernel. Indeed, the small-$q$ expansion up to $\pazocal{O}(q^2)$ is regular $\mathcal{E}_{\rm st}({\bv q}) \simeq \mathcal{E}_{\rm st}(0) - {\cal A}_2(\theta)q^2$,
where 
\begin{eqnarray}
    \mathcal{E}_{\rm st}(0)=(\lambda+2\mu)K^2 + 4\mu |K_G|
    \label{eq:Gamma0_st}
\end{eqnarray}
is the curvature-induced stretching floor, and ${\cal A}_2(\theta)$ is read from~\eqref{eq:local_kernel_general}. For ordinary elastic stability, $\mu>0$ and $\lambda+\mu>0$, it is strictly positive on a negatively curved patch. Directional information enters only through the analytic coefficient ${\cal A}_2(\theta)$ and higher powers of $q$.

The sharpest test of this mechanism is a minimal hyperbolic patch. At fixed $|K_G|$, $\mathcal{E}_{\rm st}(0)$ reaches its minimum value at $K=0$. This eliminates the mean-curvature channel and leaves only the saddle part of $K_{ab}$. Let $k_1=-k_2={\pazocal K}$, $K_G=-{\pazocal K}^2$ and $R=-2{\pazocal K}^2$, then $\mathcal{E}_{\rm st}(0)$ reduces to $4\mu{\pazocal K}^2$. The floor is therefore not a trivial ${\pazocal K}^2$ mass, surviving even when the reference surface has vanishing mean curvature.

The bending sector contribution follows by adding the normal quadratic contribution of the bending energy~\eqref{eq:Helfrich-Canham-low} for this lower-bound geometry. As was mentioned above, the fluctuation expansion involves the operator ${\cal L}_{\rm HC}$, which, for a minimal surface, is given by~\cite{PhysRevB.110.195430}, 
\begin{equation}
    {\cal L}_{\rm HC}^{\rm MinS} = \kappa(-\Delta_g+R)^2 + \sigma(-\Delta_g+R).
\end{equation}
In the same RNC local expansion, this results in
\begin{equation}
    \mathcal{E}_{\rm HC}^{\rm MinS}(q) = \kappa(q^2-2{\pazocal K}^2)^2 + \sigma(q^2-2{\pazocal K}^2),
    \label{eq:Gamma_HC_minimal}
\end{equation}
thus the total kernel considering the stretching and bending sectors is $ \mathcal{E}_{\rm total}^{\rm MinS}({\bv q})=\mathcal{E}_{\rm st}({\bv q})+ \mathcal{E}_{\rm HC}^{\rm MinS}(q)$.
Similar to~\eqref{eq:Gamma0_st}, at $q=0$,~\eqref{eq:Gamma_HC_minimal} also contributes resulting in a total lower-bound for the hyperbolic floor, as 
\begin{equation}
    \mathcal{E}_{\rm total}^{\rm MinS}(0)= 2{\pazocal K}^2 \left( 2\mu+2\kappa{\pazocal K}^2-\sigma \right).
    \label{eq:floor}
\end{equation}
Hence, the bound for positive $\mathcal{E}_{\rm total}^{\rm MinS}(0)$
is guaranteed provided $2\mu+2\kappa{\pazocal K}^2-\sigma>0$. This condition is the minimal-surface version of the curvature regularization mechanism; even after the mean curvature is set to zero, the saddle geometry produces a finite inverse normal propagator at $q=0$.

\paragraph{Structure factor and roughness saturation.}

A curvature floor directly controls long-wavelength height correlations. After integrating out in-plane phonons, the effective quadratic flexural phonon Hamiltonian becomes 
\begin{equation}
    H^{(2)}\left[\eta\right]= \frac12 \int \frac{d^{2}q}{(2\pi)^{2}}
    \left[
        \mathcal{E}_{\rm flat}(q) 
        + \mathcal{E}_{\rm curv}({\bv q})
    \right]
    |\eta({\bv q})|^2 ,
\end{equation}
$\mathcal{E}_{\rm flat}(\mathbf{q})$ is the flat-membrane flexural kernel. In the harmonic case $\mathcal{E}_{\rm flat}(\mathbf{q}) = \kappa q^4 + \sigma q^2$, with the tension term often negligible and dropped; at the standard anomalous fixed point it becomes $A q^{4-\eta_{\rm A}}$ with $A = \kappa_* q_*^{\eta_{\rm A}}$ and $\eta_{\rm A}$ the anomalous bending exponent~\cite{nelson1987fluctuations}. $\mathcal{E}_{\rm curv}(\bv{q})$ collects all curvature-induced contributions to the quadratic kernel, the stretching piece $\mathcal{E}_{\rm st}(\bv{q})$ of~\eqref{eq:local_kernel_general} together with the curvature-dependent part of $\mathcal{E}_{\rm HC}(\bv{q})$. The  static structure factor is
\begin{equation}
    S({\bv q}) \equiv \left\langle |\eta({\bv q})|^2\right\rangle = \frac{T}{A q^{4-\eta_{\rm A}} +\mathcal{E}_{\rm curv}({\bv q})}. 
\end{equation}
A finite infrared limit ${\bv q} \to 0$ is thus guaranteed at non-zero curvature. The plateau is isotropic; directional information enters only through the finite-$q$ coefficient ${\cal A}_2(\theta)$, producing curvature-selected softening at intermediate wavelengths, but not an infrared divergence. 

In particular, for minimal surfaces~\eqref{eq:floor}, the structure factor $S({\bv q})$ now converges,
\begin{equation}
    \lim_{q\to 0}S({\bv q}) = \frac{T}{\mathcal{E}_{\rm total}^{\rm MinS}(0)} .
\end{equation}
The crossover scale is defined by $\mathcal{E}_{\rm flat}(q_c)= \mathcal{E}_{\rm total}^{\rm MinS}(0)$. At the anomalous flat fixed point, 
\begin{equation}
    q_c =
    \left[
        \frac{\mathcal{E}_{\rm total}^{\rm MinS}(0)}
        {\kappa_\ast q_\ast^{\eta_{\rm A}}}
    \right]^{1/(4-\eta_{\rm A})},
\end{equation}
while for the symmetry-protected quadratic flat theory $q_c = [\mathcal{E}_{\rm total}^{\rm MinS}(0)/\kappa ]^{1/4}$. The flat-membrane scaling survives only for $q\gg q_c$, whereas for $q\ll q_c$ the response is controlled by the hyperbolic curvature floor. We write $\ell_c \equiv q_c^{-1}$ for the corresponding curvature length.

This crossover is perhaps most transparent in the local roughness, captured by the mean-squared normal displacement
\begin{equation}
    W^2(\Lambda, L) = \int_{1/L}^{\Lambda}
    \frac{d^2q}{(2\pi)^2} S({\bv q}),
\end{equation}
where $\Lambda$ is an ultraviolet cutoff. At leading isotropic order, $S({\bv q}) \simeq T/ [Aq^{4-\eta_{\rm A}}+\mathcal{E}_{\rm total}^{\rm MinS}(0)]$. Then, the thermodynamic roughness $ W_\infty^2 =\lim_{\Lambda, L\to\infty}W^2(\Lambda, L) $ is finite 
\begin{figure}[!t]
    \centering
    \includegraphics[width=\linewidth]{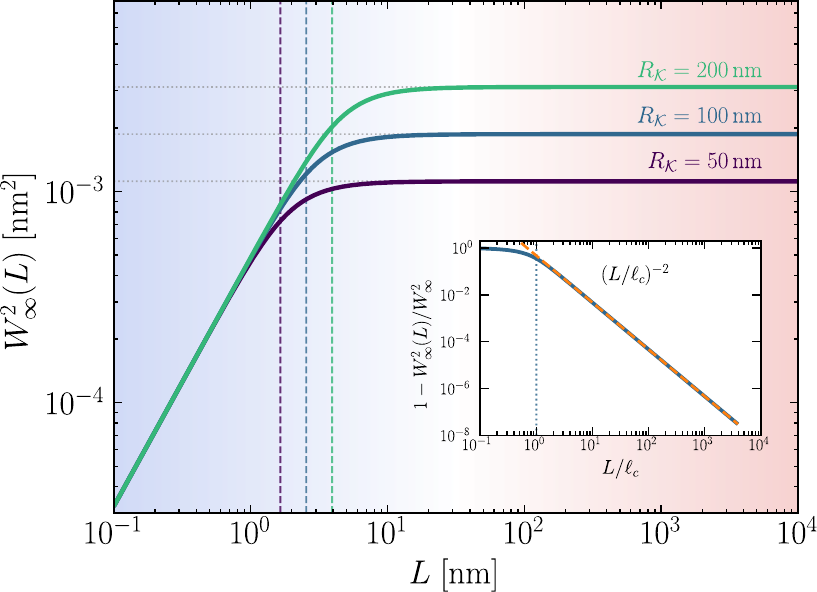}
    \caption{
    The mean-square normal displacement $W^2_{\infty}(L)$ is shown for hyperbolic curvature radii $R_{\pazocal K}=50,100,200\,{\rm nm}$ with graphene-inspired elastic parameters. Flat anomalous growth at small $L$ (blue region) is cut off at the curvature scale $\ell_c$, after which $W^2_{\infty}(L)$ saturates to a finite value $W_\infty^2$ (red region). The dotted and dashed lines mark the saturation plateaus and $\ell_c$ values, respectively. The inset shows the normalized saturation deficit; the dashed line indicates the universal large-scale approach $(L/\ell_c)^{-2}$.}
    \label{fig:roughness_saturation}
\end{figure}

\begin{align}
    W_\infty^2 &= \frac{T}{2(4-\eta_{\rm A})}
    \left(\kappa_\ast q_\ast^{\eta_{\rm A}}\right)^{-2/(4-\eta_{\rm A})} \nonumber \\
    &\hphantom{=}\;\;\, \left(\mathcal{E}_{\rm total}^{\rm MinS}(0)\right)^{-(2-\eta_{\rm A})/(4-\eta_{\rm A})}
    \csc\left[
        \frac{2\pi}{4-\eta_{\rm A}}
    \right].
    \label{eq:Winfty_eta}
\end{align}
Thus, $W^2_\infty <\infty$ follows as a direct consequence of the infrared being controlled by the curvature. In the harmonic limit $\eta_{\rm A}=0$, this reduces to $T\ell_c^2/(8\kappa)$, while with a finite ultraviolet cutoff,
\begin{equation}
    W_\infty^2(\Lambda) = \frac{T\ell_c^2}{4\pi\kappa}\,\tan^{-1}\!\left[(\Lambda\ell_c)^2\right].
\end{equation}
The approach to saturation is universal. For $L\gg \ell_c$,
the missing infrared weight is controlled only by the plateau, at leading order,
\begin{equation}
    W_\infty^2 - W_{\infty}^2(L) \simeq \frac{T \ell_c^{-2}}{4\pi {\cal E}^{\rm MinS}_{\rm total}(0)}\left(\frac{\ell_c}{L}\right)^2 ,
\end{equation}
where $W_{\infty}^2(L)=\lim_{\Lambda\to\infty}W^2(\Lambda, L)$. For $L\ll \ell_c$, the curvature floor is not yet resolved, and one recovers the anomalous benchmark flat scaling, $W^2_{\infty}(L) \sim L^{2-\eta_{\rm A}}$, or $W^2_{\infty}(L)\sim L^2$ for the harmonic tensionless membrane. Negative Gaussian curvature, therefore, converts the scale-dependent roughness of a flat crystalline membrane into a finite, curvature-controlled saturation value. 
The asymptotic behavior is shown in Fig.~\ref{fig:roughness_saturation} for elastic constants selected to represent the graphene membrane, i.e. Lam\'e parameters $\mu=910.8\,{\rm eV/nm^2}$, $\lambda=359.9~{\rm eV/nm^2}$  \footnote{ Obtained using the Young's modulus $Y_{2D}=340\,{\rm N/m}$~\cite{lee2008measurement} and Poisson's ratio $\nu=0.165$~\cite{lee2009elastic} yield $\mu=Y_{2D}/[2(1+\nu)]$ and $\lambda=Y_{2D}\nu/(1-\nu^2)$.}. The bending rigidity is $\kappa=1.44\,{\rm eV}$~\cite{wei2013bending}, and the anomalous kernel is computed using $q_*=1/a_{\rm CC}$, where $a_{\rm CC}=0.142\,{\rm nm}$ and $\eta_A =0.82$ \cite{Fasolino2007}. The inset shows the normalized saturation deficit decaying as $(L/\ell_c)^{-2}$, confirming that the plateau corresponds to the predicted infrared saturation and not to a finite-size effect. The same floor controls the orientational order whose loss defines crumpling. The normal component of~\eqref{eq:exact_strain} defines the covariant tilt $\Phi_a = \nabla_a \eta - K_{a}^{\hphantom{a}b}u_b$, which reduces to $\partial_a h$ on a flat reference, where its variance grows as $(T/2\pi \kappa) \ln (\Lambda L)$ in the harmonic flat theory. 
On the hyperbolic patch, the same computation~\cite{supp} gives a finite normal-tilt variance with curvature-cutoff logarithms involving $\ell_c$ and $|K_G|^{-1/2}$. Since the connected tilt correlator decays at large separation, $\langle{\bv n}'(x)\cdot{\bv n}'(y)\rangle$ remains proportional to the reference correlator ${\bv n}(x)\cdot{\bv n}(y)$ up to a finite fluctuation renormalization. Therefore, the thermal surface retains long-range orientational order relative to its reference. 

\paragraph{Summary.} 
We have shown that the anomalous-elasticity in thermalized crystalline membranes is preempted already at Gaussian order once the reference curvature is retained. Through minimal surfaces and local RNC truncation, the regularization reported here is a strict lower bound. The truncation discards only contributions rendered infrared-finite by the gapped in-plane propagator, and the coupling that produces the floor is fixed by rotational invariance of the embedding. These findings rest on the physical premise that thermal deformations do not originate from a flat reference. 

While the derivation is local, its implications for pristine suspended crystals are not. A corrugated membrane necessarily interleaves both signs of Gaussian curvature, and each saddle between ripples couples the flexural phonon to the in-plane shear modes, generating a mass that shields the infrared from the canonical 2D-elasticity catastrophe. On a minimal hyperbolic patch, both the height and the normal-tilt fluctuations then saturate at the curvature length rather than growing with system size. Negative Gaussian curvature thus closes the route to the crumpling instability predicted for tethered membranes---never conclusively observed---without the anharmonic interactions usually invoked to stabilize the flat phase. A well-defined flexural excitation is thereby restored throughout the infrared regime, ensuring the absence of a crumpling phase in the harmonic regime, in agreement with the suspended graphene experiment \cite{meyer2007structure}, which also shows that the system mitigates the excess area by spontaneously forming ripples.

On hyperbolic patches this branch is gapped, with long-wavelength frequency fixed by the local curvature and elastic moduli and a gapped branch carries no sound. That the flexural sector can carry sound at all in the infrared is itself a consequence of the Gaussian-order coupling: on a positively curved reference, it admits a propagating acoustic branch, a route to the flexural sound resolved by helium-atom scattering in suspended graphene~\cite{PhysRevLett.127.266102, Tomterud2025}, its spectrum and stability will be reported elsewhere. The curvature-set saturation is directly testable against spatially resolved maps of the local corrugation, and the floor's stiffening of the flexural sector should suppress the two-flexural-phonon scattering that limits carrier mobility in suspended samples.

\begin{acknowledgments}
P.A.M. would like to thank R. Kanai for supporting this research. Furthermore, P.C.-V. acknowledges the financial support provided by SNI-SECIHTI (No. CVU 92896) and financial support from the Frontiers Science grant No. CBF-2025-I-4464, financed by SECIHTI.
\end{acknowledgments}

\bibliography{references}
\bibliographystyle{apsrev4-2}


\clearpage

\appendix
\onecolumngrid

\begin{center}
{\large Curvature-Controlled Infrared Regularization of Crystalline Membranes}
\vspace{1em} \\
{\large \textbf{Supplementary Material}}
\vspace{1em} \\
Pablo A. Morales \\
    \textit{Research Division, Araya Inc., Tokyo, Japan \\
    Centre for Complexity Science, Imperial College London, London, UK}
\vspace{1em} \\
Pavel Castro-Villarreal \\
     \textit{Facultad de Ciencias en Física y Matemáticas, Benemérita Universidad Autónoma de Chiapas,\\
     Carretera Emiliano Zapata, Km. 8, Rancho San Francisco, C. P. 29050, Tuxtla Gutiérrez, Chiapas, México}
\end{center}

\renewcommand\appendixname{Supplementary Material}
\setcounter{figure}{0}
\renewcommand{\thefigure}{S\arabic{figure}}
\renewcommand{\thesection}{\arabic{section}}
\renewcommand{\theequation}{S.\arabic{equation}}
\setcounter{page}{1}

\setcounter{equation}{0}

\section{Rotational and translational invariance}
\label{app:rot_invariance}

 In this section, we show that the Green tensor (\ref{eq:exact_strain}) is unchanged under rigid motions of the $3D$ space. Let us denote by $\mathcal{R}$ a rigid spatial rotation in $3D$ space, so the embedding of the surface under a rotation becomes ${\bv X}'=\mathcal{R}{\bv X}$. For a rigid rotation, the in-plane and out-of-plane deformation fields  correspond to $u_{a}=[(\mathcal{R}-\mathbb{1}){\bv X}]\cdot {\bv e}_{a}$ and $\eta=[(\mathcal{R}-\mathbb{1}){\bv X}]\cdot{\bv n}$, respectively.  From the Weingarten-Gauss equations, follows 
 \begin{eqnarray}
    \nabla_{a}u^{c}+\eta K_{a}{}^{c}&=&\left(\mathcal{R}{\bv e}_{a}\right)\cdot{\bv e}^{c}-\delta_{a}{}^{c},\\
    \nabla_{a}\eta-K_{ac}u^{c}&=&(\mathcal{R}{\bv e}_{a})\cdot{\bv n}.
\end{eqnarray}
Note that $\mathcal{R}{\bv e}_{a}=r_{a}{}^{d}{\bv e}_{d}+r_{a}{\bv n}$ with coefficients satisfying $r_{a}{}^{c}r_{bc}+r_{a}r_{b}=g_{ab}$ due to the orthogonality property $\mathcal{R}\mathcal{R}^{T}={\bv 1}$. Substituting this into the Green tensor (\ref{eq:exact_strain}) shows $E_{ab}=0$ for any rigid rotation. For a rigid translation $\boldsymbol{X} \to \boldsymbol{X} + \boldsymbol{L}$ with constant $\boldsymbol{L} \in \mathbb{R}^{3}$, the deformation fields become $u_{a} = \boldsymbol{L} \cdot \boldsymbol{e}_{a}$ and $\eta = \boldsymbol{L} \cdot \boldsymbol{n}$. Substituting yields $E_{ab}=0$.

\section{Helfrich--Canham does not contribute to the in-plane phonon integration}
\label{app:HC-not-u}

We clarify why the in-plane phonon field $u^a$ is integrated out using
only the stretching part of the elastic Hamiltonian. The full energy is
compounded by the stretching part $H_{\rm str}$~(3) and the bending energy given by~(4), here let us here denote it $H_{\rm HC}$. The stretch energy compares the deformed metric $g'_{ab}$ with the reference metric $g_{ab}$, whereas the $H_{\rm HC}$ contribution depends only on the geometry of the deformed surface,
\begin{equation}
    H_{\rm HC}[{\bv X}'] 
    = \int_{\Sigma'} dA' \left[ \frac{\kappa}{2}(K')^2+\sigma \right],
    \label{eq:app-HHC}
\end{equation}
The distinction is important. A purely tangential deformation,
\begin{equation}
    \delta_u {\bv X} = u^a {\bv e}_a ,
    \label{eq:app-tangential-deformation}
\end{equation}
is a reparametrization of the surface in the bulk. Indeed,
${\bv X}(x+u)={\bv X}(x)+u^a\partial_a {\bv X}+\pazocal{O}(u^2)$.  Therefore, the induced geometric objects in $H_{\rm HC}$ are pulled back by a diffeomorphism. Since $H_{\rm HC}$ is reparametrization invariant, its tangential variation can only be a boundary term \cite{Capovilla_2003}. This can be checked explicitly by defining the local Helfrich--Canham density $\cal F$ as the integrand of~\eqref{eq:app-HHC}. Under~\eqref{eq:app-tangential-deformation},
\begin{equation}
    \delta_u g_{ab}
    = \pounds_u g_{ab} = \nabla_a u_b+\nabla_b u_a ,
    \qquad
    \delta_u\sqrt{g} = \sqrt{g}\,\nabla_a u^a ,
\end{equation}
while $K$ is a scalar on the surface, so
\begin{equation}
    \delta_u K = u^a\nabla_a K .
\end{equation}
Hence
\begin{align}
    \delta_u H_{\rm HC}
    &= \int dA \left[(\nabla_a u^a){\cal F}
    + \kappa K u^a\nabla_a K
    \right] \nonumber \\
    &= \int dA \left[(\nabla_a u^a){\cal F} +
    u^a\nabla_a{\cal F} \right] \nonumber \\
    &= \int dA\,\nabla_a(u^a{\cal F}) .
\end{align}
Thus, for a closed surface or for boundary conditions that fix the
tangential boundary variation, $\delta_u H_{\rm HC}=0$ in the bulk. Equivalently, although $K'$ may contain $u^a$ when expanded in fixed material coordinates, this dependence is a pure reparametrization of the functional. It does not modify the bulk Gaussian integration over $u^a$.

Although $H_{\rm HC}$ does not modify the bulk Gaussian integration over the tangential phonons, it must still be kept in the normal fluctuation sector. For a normal deformation $\delta_\perp {\bv X}=\eta {\bv n}$, we use
\begin{equation}
    \delta_\perp \sqrt{g} = \sqrt{g} K\eta ,
    \qquad
    \delta_\perp K= -\Delta\eta-K_{ab}K^{ab}\eta ,
\end{equation}
The first normal variation is therefore
\begin{equation}
    \delta_\perp H_{\rm HC}
    = \int dA \left[- \kappa K\Delta\eta - \kappa K K_{ab}K^{ab}\eta + \frac{\kappa}{2}K^3\eta + \sigma K\eta \right].
\end{equation}
After integrating the Laplacian term by parts, assuming a closed surface
or boundary conditions that remove boundary contributions, one obtains
\begin{equation}
    \delta_\perp H_{\rm HC}
    =\int dA\, \eta\left[-\kappa\Delta K
    -\kappa K K_{ab}K^{ab} +\frac{\kappa}{2}K^3 +\sigma K \right] + \text{boundary}.
\end{equation}
Using the Gauss-Codazzi relation, this may be rewritten as
\begin{equation}
    \delta_\perp H_{\rm HC}
    =\int dA\, \eta\left[-\kappa\Delta
    -\kappa R +\frac{\kappa}{2}K^2 +\sigma \right]\!K + \text{boundary}.
    \label{eq:app-HC-normal-var}
\end{equation}
This expression shows explicitly that the bulk HC variation is controlled by the normal displacement. The tangential field $u^a$ contributes only through reparametrizations of the surfaces and therefore does not enter the bulk phonon Gaussian.
If the reference surface is an equilibrium shape of the HC functional, the bracket at Eq.~\eqref{eq:app-HC-normal-var} vanishes. The fluctuation expansion then starts at quadratic order,
\begin{equation}
    H_{\rm HC}[{\bv X}^{\prime}]
    = H_{\rm HC}[{\bv X}]
    + \frac{1}{2}
    \int dA\,\eta\,{\cal L}_{\rm HC}\,\eta +
    \cdots,
\end{equation}
where ${\cal L}_{\rm HC}$ is a fourth-order geometric operator. For minimal surfaces this operator is given ${\cal L}_{\rm HC}^{\rm MinS} = \kappa(-\Delta_g+R)^2 + \sigma(-\Delta_g+R)$~\cite{PhysRevB.110.195430}.

\section{Saturation of normal-tilt fluctuations and absence of crumpling}
\label{app:tilt}

Crumpling is the loss of long-range orientational order of the surface
normals. On a flat reference, this is diagnosed by $\langle|\partial_a h|^2\rangle$ with $h$ being the height function, which diverges logarithmically with system size in the harmonic theory. On a curved reference, the covariant tilt field is fixed by the geometry; from~(1), the normal component of the deformed tangent vector defines
\begin{equation}
    \Phi_a \equiv \nabla_a\eta - K_a{}^{b}u_b .
    \label{eq:tilt}
\end{equation}
To find the deformed unit normal, choose at a point an oriented orthonormal frame ${{\bv e}_1,{\bv e}_2,{\bv n}}$ and write the deformed tangents as ${\bv e}'_a=A_a{}^b {\bv e}_b+\Phi_a {\bv n}$, with $A_a{}^b=\delta_a{}^b+\nabla_a u^b+\eta K_a{}^b$. The area normal is

\begin{equation}
    {\bf N}'\equiv {\bf e}'_1\times {\bf e}'_2=(\det A)\sqrt{g}({\bf n}-\psi^a {\bf e}_a),\qquad A_a{}^b\psi_b=\Phi_a.
\end{equation}
The unit normal is therefore ${\bv n}'=\frac{{\bf N}'}{|{\bf N}'|}=\frac{{\bv n}-\psi^a {\bv e}_a}{\sqrt{1+\psi^2}}$
. Since $\psi_a=\Phi_a+\pazocal{O}(2)$, the Gaussian-order normal rotation is
\begin{equation}
    {\bv n}'\simeq \left(1-\frac{1}{2}\Phi_a\Phi^a\right){\bv n}-\Phi^a {\bv e}_a 
    \label{eq:app-unit_normal_exp}
\end{equation}
The field~\eqref{eq:tilt} enters the Green strain~(2) as a complete square, reduces to $\partial_a h$ in the flat limit, and transforms as $({\cal R}{\bv e}_a)\cdot {\bv n}$ under rigid rotations (see SM: Rotational invariance), 
as a normal tilt must.
The membrane retains the orientational order of its reference shape if
$\langle\Phi_a\Phi^a\rangle$ remains bounded as $L\to\infty$ and
tilt correlations decay.

In the local frame of~(17), write $\Theta_{ab} \coloneqq \lambda K\delta_{ab}+2\mu K_{ab}$, so that the source reads $J_b(\bv q)=-i\Theta_{b}^{\hphantom{b}a}q_a \eta(\bv q)$, and write the in-plane propagator~(14) as $G=g_T P^{T}+g_L P^{L}$ with $g_T=[\mu(q^2+|K_G|)]^{-1}$ and $g_L=[(2\mu+\lambda)q^2+\mu|K_G|]^{-1}$. Completing the square,
$u_a=\tilde u_a-(GJ)_a$ with $\langle\tilde u_a (\bv q)\tilde u_b(-\bv{q})\rangle = T G_{ab}(\bv q)$ and $\langle\tilde u_a(\bv q)\eta(-\bv{q})\rangle=0$, gives
\begin{equation}
    \Phi_a(\bv q) = i V_a(\bv q) \eta(\bv q) - K_{ab} \tilde u_b(\bv q),
    \qquad
    V_a \equiv q_a - K_a{}^b G_b{}^c \Theta_c{}^d q_d .
\end{equation}
In the $(\bv n,\bv t)$ basis, with $k_t \coloneqq K-k_n$,
\begin{align}
    V_n &= q\big[1 - g_L k_n(\lambda K+2\mu k_n) - 2\mu g_T \tau_g^2 \big], \\
    V_t &= - q \tau_g \big[g_L(\lambda K+2\mu k_n) + 2\mu g_T k_t \big].
\end{align}
With $\langle|\eta|^2\rangle = S(q)=T/\mathcal{E}_{\rm total}(q)$, the
tilt correlator is
\begin{equation}
    C_{ab}(\bv q) = V_a V_b S(q) + T\big[g_L (K\bv n)_a (K\bv n)_b + g_T (K\bv t)_a (K\bv t)_b \big],
\label{eq:Cab}
\end{equation}
which reduces to $q_a q_b S(q)$ in the flat limit. On the minimal hyperbolic patch, $K=0$, $k_1=-k_2=\pazocal{K}$, $K_G=-\pazocal{K}^{2}$, the traceless $K_{ab}$ obeys $K_{a}{}^{c}K_{cb}=\pazocal{K}^{2}\delta_{ab}$. As $q\to0$ both channels coincide, $g_L,g_T\to(\mu \pazocal{K}^{2})^{-1}$, so $V_t\to0$ and $V_n\to-q$ so the angular structure cancels, $|V|^2\to q^{2}$ isotropically, and the term proportional to $\eta$ vanishes at $q=0$. The zero-wavevector tilt density is carried by the in-plane sector,
\begin{equation}
    C_{ab}(\bv q\to 0) = \frac{T}{\mu} \delta_{ab},
\end{equation}
finite and isotropic. The local variance $\langle\Phi_a\Phi^a\rangle = \int\!\tfrac{d^2 q}{(2\pi)^2} C_{a}^{\hphantom{a}a}(\bv q)$ splits accordingly,
\begin{align}
    \langle\Phi^2\rangle_{\tilde u}
    &= \frac{T \pazocal{K}^{2}}{4\pi}\bigg[\frac{1}{\mu}
    \ln\Big(1+\frac{\Lambda^{2}}{\pazocal{K}^{2}}\Big) + \frac{1}{2\mu+\lambda}
    \ln\Big(1+\frac{(2\mu+\lambda)\Lambda^{2}}{\mu \pazocal{K}^{2}}\Big)\bigg],\\
    \langle\Phi^2\rangle_{\eta}
    &= \frac{T}{8\pi\kappa}\,\ln\!\big(1+\Lambda^{4}\ell_c^{4}\big)
    + \Delta,
\end{align}
where $\Delta=\int\!\tfrac{d^2q}{(2\pi)^2} (|V|^{2}-q^{2}) S(q)$ is
finite at both ends, $|V|^{2}-q^{2}=\pazocal{O}(q^{2})$ with $S$ saturated in the infrared, while $|V|^{2}-q^{2}=\pazocal{O}(\pazocal{K}^{2})$ with $S\sim T/\kappa q^{4}$ in the ultraviolet. The flat-membrane result $\langle|\partial h|^{2}\rangle = (T/2\pi\kappa)\ln(\Lambda L)$ is thus reproduced with the system size replaced by the curvature lengths, $L\to\ell_c$ in the flexural channel and $L\to\ell_G\equiv \pazocal{K}^{-1}$ in the shear channel. The crumpling logarithm is cut by geometry, and $\lim_{L\to\infty} \langle\Phi^2\rangle$ is finite at fixed temperature. The remaining logarithmic divergence is ultraviolet, cut at the lattice scale exactly as on a flat membrane, and does not affect the infrared statement.

In real space, the $\tilde u$ contribution decays as massive 2D Green functions, while the flexural contribution is controlled by the poles of $S(q)$. Therefore the connected tilt correlator
\begin{equation}
    C_\Phi(x,y)\equiv \langle\Phi^a(x)\Phi^b(y)\rangle {\bv e}_a(x)\cdot {\bv e}_b(y)
    \label{eq:app-C_Phi}
\end{equation}
decays beyond the curvature-controlled lengths. Using~\eqref{eq:app-unit_normal_exp} the normal-normal correlator is
\begin{equation}
    \langle {\bv n}'(x) \cdot {\bv n}'(y)\rangle=
    \left[1-\frac12\langle\Phi^2(x)\rangle-\frac12 \langle\Phi^2(y)\rangle\right]{\bv n}(x)\cdot{\bv n}(y) + \langle \Phi^a(y)\Phi^b(y)\rangle {\bv e}_{a}(x)\cdot{\bv e}_{b}(y) + \pazocal{O}(\Phi^3).
\end{equation}
where the linear terms vanish for fluctuations centered around the reference surface. In a locally homogeneous patch, $\langle \Phi^2(x) \rangle =\langle\Phi^2(y)\rangle=\langle\Phi^2\rangle$, and since $C_\Phi(x,y)\to0$ at large separation,
\begin{equation}
    \langle {\bv n}'(x)\cdot {\bv n}'(y)\rangle\to \left[1-\langle\Phi^2\rangle\right]{\bv n}(x)\cdot {\bv n}(y).
    \label{eq:app_normal_correlator_asymp}
\end{equation}
The omitted connected term in~\eqref{eq:app_normal_correlator_asymp} is precisely the decaying tilt correlator~\eqref{eq:app-C_Phi}. At coincident points it restores unit normalization, since $\langle \Phi^a(x) \Phi^b(x) \rangle g_{ab}=\langle\Phi^2\rangle$. Thus the thermal normal field remains locked to the reference normal field with a finite fluctuation renormalization. Crumpling, understood as unbounded growth of normal-tilt fluctuations with system size, is absent because $\langle\Phi^2\rangle$ is infrared finite.

\end{document}